\documentclass[
 aip,
 amsmath,amssymb,
 reprint,
]{revtex4-1}

\usepackage{graphicx}
\usepackage{dcolumn}
\usepackage{bm}
\usepackage{color}

\usepackage[utf8]{inputenc}
\usepackage[T1]{fontenc}
\usepackage{mathptmx}
\usepackage{natbib}

\begin{document}

\preprint{AIP/123-QED}

\title{A Simple Numerical Method for Evaluating Heat Dissipation from Curved Wires with Periodic Applied Heating}
 
\author{Gabriel R. Jaffe}
\email{gjaffe@wisc.edu}
\author{Victor W. Brar}

\affiliation{Department of Physics, University of Wisconsin-Madison, Madison, Wisconsin 53706, USA}

\author{Max G. Lagally}
\affiliation{Department of Materials Science and Engineering, University of Wisconsin-Madison, Madison, Wisconsin 53706, USA}

\author{Mark A. Eriksson}
\affiliation{Department of Physics, University of Wisconsin-Madison, Madison, Wisconsin 53706, USA}

\date{\today}

\begin{abstract} 
In many situations, the dual-purpose heater/thermometer wires used in the three-omega method -- one of the most precise and sensitive techniques for measuring the thermal conductivity of thin films and interfaces -- must include bends and curves to avoid obstructions on the surface of a sample.  Although the three-omega analysis assumes that the heating wire is infinitely long and straight, recent experimental work has demonstrated that in some cases curved-wire geometries can be used without introducing detectable systematic error.  We describe a general numerical method that can be used to calculate the temperature of three-omega heating wires with arbitrary wire geometries.  This method provides experimentalists with a simple quantitative procedure for calculating how large the systematic error caused by a particular wire asymmetry will be.  We show calculations of two useful cases: a straight wire with a single bend of arbitrary angle and a wire that forms a circle.  We find that the amplitude of the in-phase temperature oscillations near a wire that forms a circle differs from the prediction using the analytic straight-line source solution by $<$12\%, provided that the thermal penetration depth is less than ten times the radius of curvature of the wire path. The in-phase temperature amplitude 1.5 wire widths away from a 90$^{\circ}$ bend in a wire is within 11\% of the straight-line source prediction for all penetration depths greater than the wire width.  Our calculations indicate that the straight-line source solution breaks down significantly when the wire bend angle is less than 45$^{\circ}$.  
\end{abstract}

\maketitle

Precise measurements of the thermal conductivity of thin films and the interfaces between them are crucial for furthering our understanding of nanoscale thermal transport.\cite{Cahill_JAP_2003,Pop_NR_2010}   One of the most reliable and widely used techniques to measure thermal conductivity, the three-omega method, involves periodic Joule heating of a metal wire fabricated on the material of interest.\cite{Cahill_RSI_1990,Lee_JAP_1997,Kim_APL_2000,Kim_PRL_2006,Chen_APL_2009,Schroeder_PRL_2015,Yang_AFM_2018,Jaffe_ACSAMI_2019,Xu_JAP_2019,Velarde_ACEAMI_2019} Heating frequencies are chosen such that the thermal wave emitted by the wire penetrates only a few hundred microns into the substrate before dissipating.  The rapid dissipation of the thermal wave confines the volume of the sample being probed by the experiment, thereby increasing the measurement's sensitivity to thermal signals from thin films and interfaces near the sample's surface, while simultaneously minimizing the effects of convective and radiative heat loss.\cite{Cahill_PRB_1987}  In order to extract a measure of thermal conductivity from the temperature fluctuations in the wire, the three-omega method relies on a solution to the heat equation that assumes the heating wire is infinitely-long and straight.  In practice, it can be challenging to achieve a perfectly straight heating wire on a real sample because the heating wires must often be routed around obstacles such as residue, bubbles, and wrinkles to reach the area of interest.\cite{Cavallo_SoftMat_2010,Rogers_NAT_2011,Pizzocchero_NC_2016}   The sections of heating wire that do not lie on a straight line may introduce systematic errors into the determination of thermal conductivity that are difficult to predict. 

A recent experimental work has demonstrated that, to some degree, the straight wire rule can be bent: thermal conductivity measurements of SiO$_2$ thin films on Si substrates performed with both straight and curved heating wires were found to agree to within the experimental uncertainty.\cite{Jaffe_APL_2020}  This study provided a lower bound for how curved a heating wire can be without noticeably affecting the accuracy of a measurement.  What is currently not known is at what point bends or curves in a wire cause significant deviations in the wire temperature from the straight-wire prediction, and how these deviations scale with parameters such as wire bend angle, wire curvature, and the thermal penetration depth of the applied thermal wave.

\begin{figure*}[t]
	\centering
	\includegraphics[width=1.5\columnwidth]{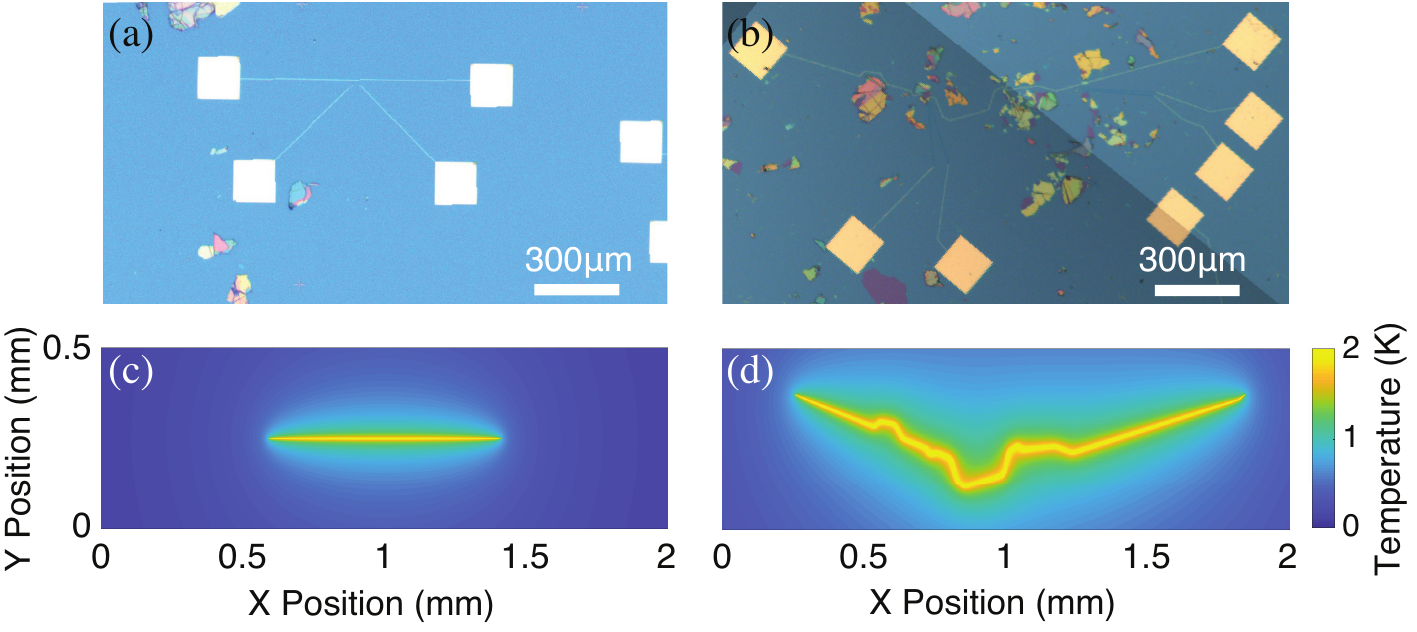}
	\caption{(a) An optical image of a gold 3$\omega$ heating wire with voltage probes fabricated on a SiO$_2$/Si substrate where the wire is perfectly straight. (b) An optical image of a heating wire that includes many bends as it navigates around obstacles on the sample surface in order to pass over exfoliated flakes of hexagonal boron nitride.  The plots (c) and (d) show calculations using the point source superposition method of the amplitude of the in-phase temperature oscillations about the steady state temperature for periodic heating of the wires shown in (a) and (b), respectively.}
	\label{fig:1}
\end{figure*}

In this Letter, we show that superimposing the time dependent temperature profile of many point sources of heat allows us to approximate the periodic heating of a wire with arbitrary shape.  Importantly, we show that an analytical solution for the time-dependent heat flow from a point source into a semi-infinite solid can be used, making this procedure both numerically simple and fast. This method can be employed to test parameters in a three-omega experiment, regardless of the heating wire geometry. These calculations assume all thermal transport occurs through heat conduction in the substrate and that thermal radiation and convection are negligible, as is the case in three-omega experiments.\cite{Cahill_RSI_1990}  Using this technique, we perform calculations of wires that bend or curve and compare the temperature fluctuations near these wires to the analytic heat equation solution for an infinitely long and straight wire. The geometries evaluated in this work provide a quantitative prediction of the systematic error introduced into a three-omega measurement when the heating wire used for the experiment includes a single bend or forms a circle.  Our numerical approach enables us to examine in detail how parameters such as thermal penetration depth, wire bend angle, distance to the bend in the wire, and wire radius of curvature affect the temperature rise near the wire.  We find that when the heating wire forms a circle, the in-phase temperature amplitude near the wire is within 12\% of the straight-line source prediction when the thermal penetration depth is less than ten times the wire radius of curvature.  The straight-line source solution under predicts the temperature 1.5 wire widths away from a 90$^{\circ}$ bend in a wire by $<$11\% for all penetration depths greater than the wire width. 

In addition to being useful in determining the thermal transport properties of semi-infinite substrate. the technique here is also useful for three-omega experiments where the thermal resistance of a thin film or interface directly beneath the heating wire is of interest.  For such an experiment, the thermal resistance of the thin film or interface is determined by subtracting the expected thermal resistance contribution of the supporting substrate.\cite{Cahill_PRB_1994} The method described here builds upon the standard three omega analysis by enabling a prediction of the substrate contribution for a wire of arbitrary shape.

\begin{figure}[t]
	\centering
	\includegraphics[width=1.0\columnwidth]{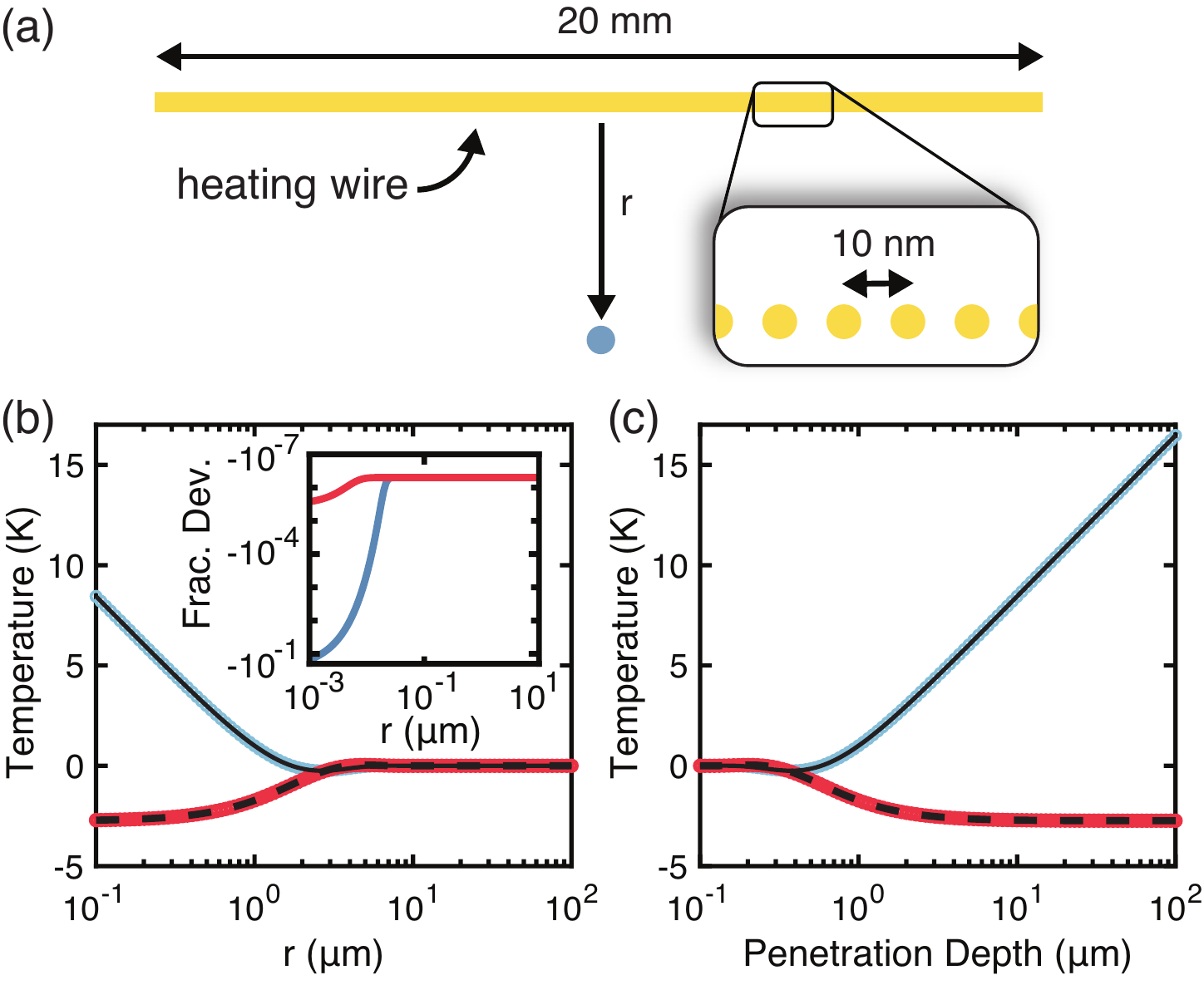}
	\caption{(a) A schematic diagram of a heating wire approximated by a linear array of point heat sources.  The temperature is evaluated at the blue point, a distance $r$ from the wire.  (b) A comparison of the in-phase (solid line) and 90 degree out-of-phase (dashed line) temperature amplitude as a function of distance $r$ from an infinite-line source of heat with sinusoidal time dependence as calculated by equation 1 from Ref.\,\citenum{Cahill_RSI_1990} to the amplitude calculated by a linear superposition of the point-source solution given here in equation (\ref{eq:1}) (blue and red circles). The thermal penetration depth for both calculations is 1\,$\mu$m. The inset shows the fractional deviation between the two methods as a function of $r$.  (c) A comparison of the infinite-line source solution with the superposition method as a function of penetration depth at fixed distance of $r$ = 1\,$\mu$m away from the wire.}
	\label{fig:2}
\end{figure}

An optical image of a straight three-omega heating wire is shown in Fig.\,\ref{fig:1}(a), and a wire that includes many bends in order to navigate around surface features is shown in Fig.\,\ref{fig:1}(b).  The amplitude of the temperature oscillations in phase with the applied ac heating to these wires is calculated using the superposition method and shown in Fig.\,\ref{fig:1}(c) and (d).  For this calculation, the heating wires are represented by point heat sources spaced by 10\,nm.  The temperature at each position in Fig.\,\ref{fig:1}(c) and (d) is the sum of the temperature rise given by Eqn.\,(\ref{eq:1}) below over all point sources, where $r$ is the distance to each heat source. These calculations do not require significant computational resources to run; both of the calculations shown in Fig.\,\ref{fig:1} were evaluated in a few minutes on a single processing core.  Our methodology is based on superimposing a dense array of ac point-heat sources in order to approximate the ac temperature fluctuations from a given wire geometry as seen in Fig.\,\ref{fig:2}(a).  In equation (\ref{eq:1}), we provide the solution to the heat equation for the temperature oscillations about the steady state temperature for a point source of heat with heat flux sinusoidal in time at the origin of an infinite solid. The mathematical techniques for solving the heat equation for high-symmetry cases are well documented;\cite{Carslaw_Book_1947} however, the solution for this specific case is not readily available in the literature.  We therefore provide a derivation of this solution in the supporting information.  The temperature oscillations $T$ about the steady state temperature at time $t$ a distance $r$ away from the point-heat source emitting $Q$ units of heat per unit time at frequency $\omega$ into an semi-infinite medium with thermal diffusivity $D$ and thermal conductivity $\kappa$ are given by
\begin{equation}\label{eq:1}
	T(r,t) = \frac{Q }{2\pi\kappa} \frac{e^{-\sqrt{\frac{\omega}{2D}}r }}{r} \sin \left( \omega t - \sqrt{\frac{\omega}{2D}}r \right).
\end{equation}
%

\begin{figure*}[t]
	\centering
	\includegraphics[width=1.6\columnwidth]{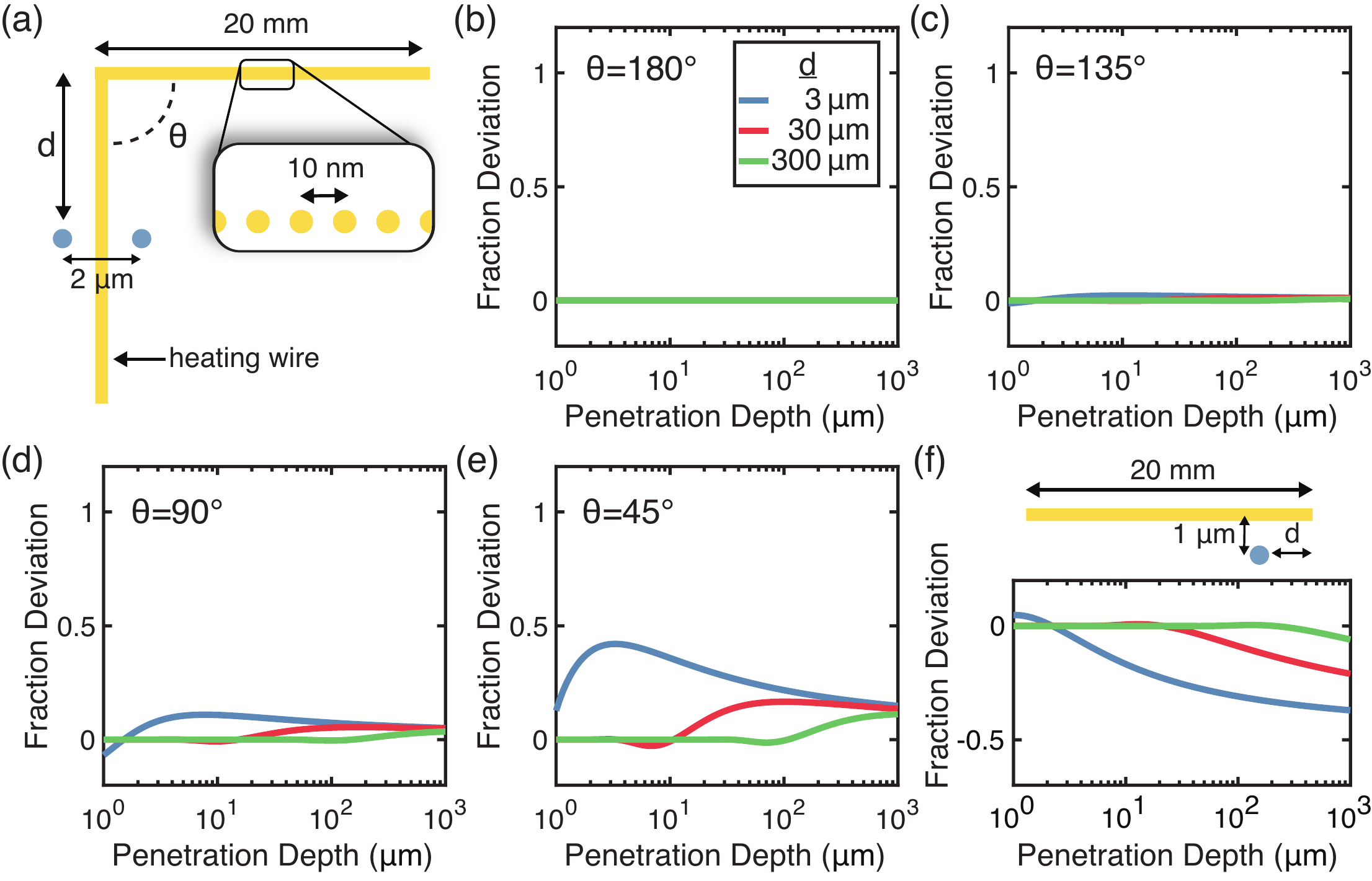}		
	\caption{(a) Schematic diagram of a wire with a bend angle of $\theta = 90^{\circ}$.  Periodic heating of the wire is approximated by a linear array of ac point heat sources spaced 10\,nm apart. The temperature amplitude in phase with the applied heating is evaluated and averaged between the blue points located 1\,$\mu$m on either side of the wire and a distance $d$ away from the bend in the wire.  (b-e) The fractional deviation of the in-phase temperature amplitude calculated using the superposition method of a wire with bend angle $\theta$ from that of an infinite line as a function of penetration depth for three different distances to the bend $d$. (f) The fractional deviation from the infinite-line source solution a distance $d$ from the end of a 20\,mm long wire calculated using the superposition method.}
	\label{fig:3}
\end{figure*}

We verify our superposition method by comparing the temperature profile calculated from the superposition of equation (\ref{eq:1}) for a dense linear array of point sources to the exact solution for the temperature profile from an infinite-line source of heat.\cite{Cahill_RSI_1990} In order to approximate an infinitely long wire, the temperature was evaluated at the center of a 20\,mm long array of point heat sources as seen in Fig.\,\ref{fig:2}(a).  For all calculations shown here, we assume $D =$ 10$^{-4}$\,m$^2$\,s$^{-1}$ and $\kappa =$ 100\,W\,m$^{-1}$\,K$^{-1}$. In Fig.\,\ref{fig:2} (b) and (c) we show the amplitude of the temperature oscillations that are in phase (blue points) and ninety degrees out of phase (red points) with the applied heating as the distance $r$ to the wire and heating frequency are varied.  The exact solutions for the in-phase and out-of-phase temperature amplitudes for an infinite wire are shown as lines and agree well with the superposition method. The inset of Fig.\,\ref{fig:2}(b) shows that the fractional deviation between the temperature calculated using the superposition method and the infinite-line source solution is less than 10$^{-6}$ as long as $r$ is greater than ten times the point heat source spacing. Physically, increasing the density of point sources produces a result closer to a continuous wire and the fractional deviations in Fig.\,\ref{fig:2}(b) reflect the spacing we have chosen.  Note that the horizontal axis in Fig.\,\ref{fig:2}(c) is the thermal penetration depth $\lambda = \sqrt{D/\omega}$.  This is the important physical quantity for ac thermal measurements.  As an example, a line heat source at 1\,kHz for thermal transport in the cross-plane direction of bulk hexagonal boron nitride would have $\lambda\sim$21$\mu$m.\cite{Jaffe_ARXIV_2021}  In Fig.\,\ref{fig:2} (b) and (c) we see that the point source superposition method correctly predicts that the temperature near an infinite line source is proportional to $\lambda/r$, despite the fact that the point source solution given in equation (\ref{eq:1}) is not.

\begin{figure}[t]
	\centering
	\includegraphics[width=1.0\columnwidth]{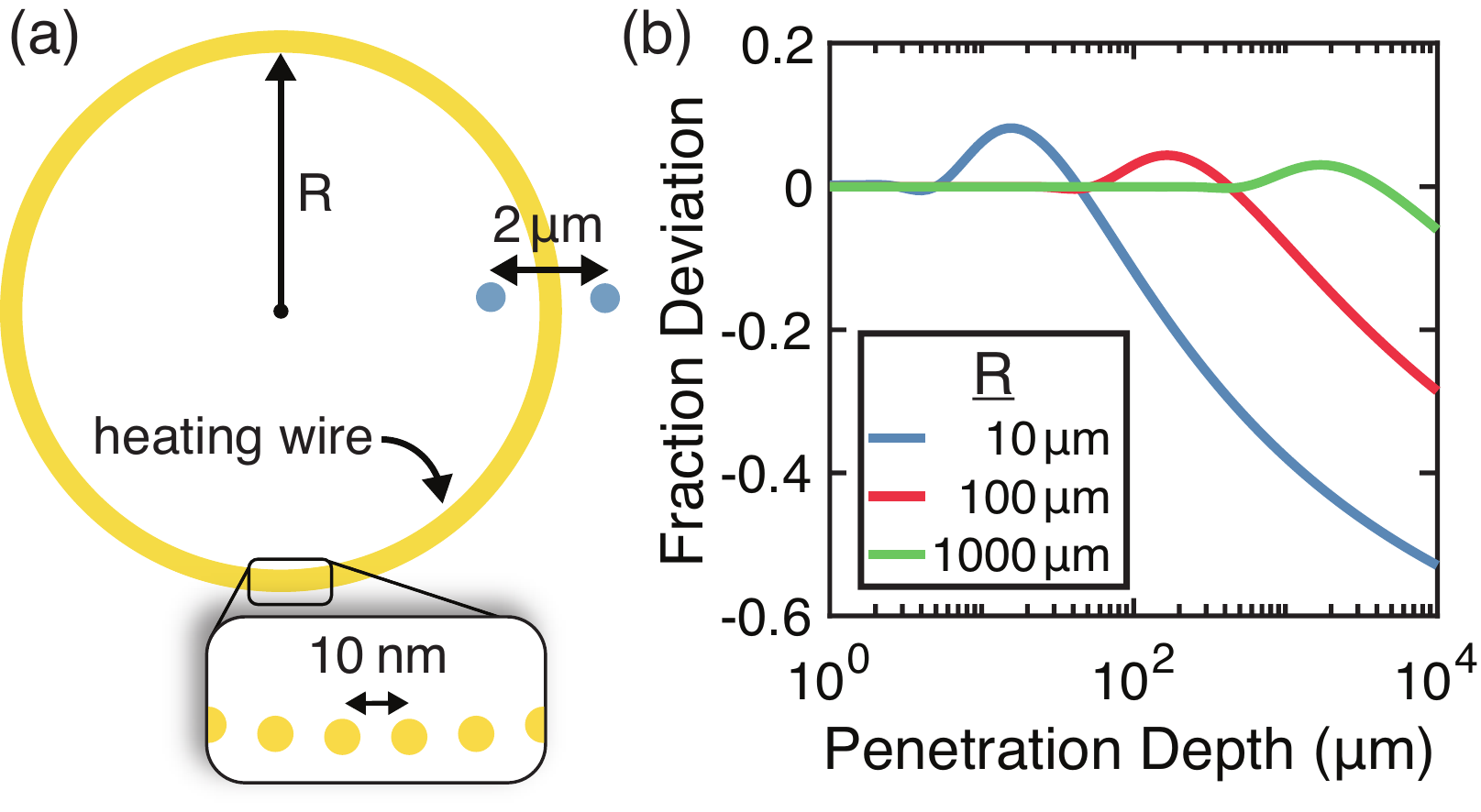}
	\caption{ (a) A schematic diagram of a wire that forms a circle of radius R. The temperature is averaged at the blue points. (b) The fractional deviation from the infinite-line source solution of the in- phase temperature amplitude for three different radii of curvature as a function of penetration depth.}
	\label{fig:4}
\end{figure}

In order to quantify how the amplitude of the in-phase temperature oscillations calculated using the superposition method  differs from the infinite-line source solution, we define the fractional deviation parameter $\epsilon$ as
\begin{equation}\label{eq:2}
	\epsilon = \frac{T_{superposition}}{T_{line}}-1,
\end{equation}
where $T_{superposition}$ is the temperature calculated using the superposition method and $T_{line}$ is the temperature calculated from the infinite-line source solution.  When $\epsilon \approx 0$, it means that the temperature near a wire is well described by the  infinite-line source heat equation solution.  For all other values of $\epsilon$, the infinite-line source solution either under- or overestimates the wire temperature by a factor of $\epsilon$.  We explore how $\epsilon$ changes for two simple cases that characterize many features of many real-world asymmetric wire geometries: a straight wire with a single bend of angle $\theta$, and a wire that forms a circle.  These calculations provide a practical guide for experimentalists to predict the systematic error introduced by asymmetric wire geometries when using the three-omega method.   We calculate $\epsilon$ as a function $\lambda$ which itself is proportional to the heating frequency of the wire.


In Fig.\,\ref{fig:3} we show the fractional deviation $\epsilon$ of the in-phase temperature amplitude near wires that contain a single bend of angle $\theta$.  In order to approximate the in-phase temperature amplitude of a wire with a finite width of 2\,$\mu$m, we evaluate the temperature at a fixed offset of 1\,$\mu$m on either side of the point sources (blue points in Fig.\,\ref{fig:3}(a)).  A fixed offset from the point sources is required because the point and line source heat equation solutions diverge as $r$ approaches zero.  The temperature is evaluated at three different distances $d$ from the bend in the wire in order to show how proximity to the bend changes $\epsilon$.  In Fig.\,\ref{fig:3}(b-e) we plot the fractional deviation $\epsilon$ for wires with bend angles ranging from $\theta = 180^{\circ}$ to $\theta = 45^{\circ}$ as a function of thermal penetration depth.  In the case of $\theta = 180^{\circ}$ the wire is perfectly straight and $\epsilon$ is always zero, as expected. 

We find that the maximum $\epsilon$ for a given bend angle occurs when the distance from where the temperature is evaluated to the bend in the wire, $d$, is the same as the thermal penetration depth.  We note that in equation (\ref{eq:1}), the thermal wave emanating from each point source picks up a phase $\phi = \frac{r}{\sqrt{2}\lambda}$ at distance $r$ for penetration depth $\lambda$.  The effect of the bend in the wire on $\epsilon$ should therefore be greatest when the angular frequency satisfies $\lambda(\omega) \approx d$.  In Fig.\,\ref{fig:3}(b-e), when  $\lambda > d$, $\epsilon$ decreases with increasing penetration depth.  We find that for bent wires with $\theta \geq$ 90$^{\circ}$, the fractional deviation remains less than 11\% for all penetration depths, even when the temperature is measured 1.5 wire widths from the bend.  Significant values of $\epsilon$ appear to occur only for more acute angles $\theta<45^{\circ}$.  These results will scale with wire width and penetration depth. For example, the fractional deviation for a 2\,$\mu$m wide wire at a penetration depth of 10\,$\mu$m will be the same for a 100\,$\mu$m wide wire at a penetration depth of 500\,$\mu$m. 

We now turn to the question of nonidealities that occur near the end of a finite length wire. Numerical simulations of three-omega experiments have shown that the heat dissipates three-dimensionally from the ends of a wire of finite length, which reduces the temperature of the wire below the infinite-line prediction.\cite{Jacquot_JAP_2002} We observe this effect in Fig.\,\ref{fig:3}(f), where we find that $\epsilon < 0$ when $\lambda > d$, where $d$ is the distance to the end of a wire of finite length.

The fractional deviation of the temperature of heating wires that form a circle with radius $R$ from the infinite-line source solution is shown in Fig.\,\ref{fig:4}.  For all calculations, the in-phase temperature amplitude is averaged at points offset 1\,$\mu$m on either side of the circle to approximate a wire with finite width of 2\,$\mu$m.  We find that the fractional deviation remains small for a wide range of penetration depths, with $\lvert \epsilon \rvert <$ 0.12 for penetration depths $\lambda< 10\,R$. When $\lambda << R$, the wire appears infinitely long and straight relative to the thermal wave, causing $\epsilon$ to approach 0.  When $\lambda >> R$, the wire acts like a point source of heat rather than an infinite line, and $\epsilon$ approaches $-1$. We observe small peaks in $\epsilon$ when the penetration depth is slightly larger than the wire radius of curvature.

These calculations indicate that there is almost always a practical range of penetration depths, and therefore heating frequencies, that can be chosen for a three-omega measurement such that asymmetric wire geometries see nearly the same temperature rise as infinite straight wires.  We find that the in-phase temperature amplitude near a periodically heated wire that forms a circle differs from that of an infinite straight wire by $<$\,12\% provided that the thermal penetration depth is smaller than ten times the wire radius.  Wires with $>$\,90$^{\circ}$ bends see small fractional deviations from the infinite-line source solution at all penetration depths provided that the temperature of the wire is measured at least a few wire widths away from the bend.  Care must be taken when performing experiments with wires that have acute angles $<$\,45$^{\circ}$ because the temperature near the bends in these wires can be more than 40\% above what is predicted from the infinite-line source solution, particularly when the distance between the position where the temperature of the wire is measured to the wire bend is equal to the penetration depth.  This point-source superposition method provides experimentalists with a useful and computationally inexpensive tool for evaluating different heating wire geometries for three-omega experiments.

\vspace{-15pt}
\section*{Supplementary Material}
\vspace{-10pt}
See supplementary material for the derivation of the heat equation solution given in Eqn.\,(\ref{eq:1}) and more detailed numerical calculations of the effects of the finite width of the wire.

\vspace{-15pt}
\section*{Acknowledgements}
\vspace{-10pt}

Calculations are performed with support from US DOE Basic Energy Sciences DE-FG02-03ER46028. 

The data that support the findings of this study are available from the corresponding author upon reasonable request.

\bibliographystyle{apsrev4-1}
\bibliography{./Bibliography}

\end{document}